\documentclass[a4paper,11pt]{elsarticle}

\usepackage[utf8x]{inputenc}
\usepackage{amsmath}
\usepackage{graphicx}
\usepackage{xcolor}

\begin{document}
\begin{frontmatter}
\title{\textbf{The long duration cryogenic system of the OLIMPO balloon--borne experiment: design and in--flight performance}}
\author[Rome,mysecondaryaddress]{\textbf{A. Coppolecchia} \corref{mycorrespondingauthor}}
\ead{alessandro.coppolecchia@roma1.infn.it}
\author[Rome,mysecondaryaddress]{\textbf{L. Lamagna}}
\author[Rome,mysecondaryaddress]{\textbf{S. Masi}}
\author[Cardiff]{\textbf{P.A.R. Ade}}
\author[Rome]{\textbf{\\G. Amico}}
\author[Rome,mysecondaryaddress]{\textbf{E.S. Battistelli}}
\author[Rome,mysecondaryaddress]{\textbf{P. de Bernardis}}
\author[Rome,mysecondaryaddress]{\textbf{F. Columbro}}
\author[Rome,ESA]{\textbf{L. Conversi}}
\author[Rome,mysecondaryaddress]{\textbf{G. D'Alessandro}}
\author[Rome,mysecondaryaddress]{\textbf{M. De Petris}}
\author[Milano]{\textbf{M. Gervasi}}
\author[Rome,Milano]{\textbf{F. Nati}}
\author[Rome]{\textbf{L. Nati}}
\author[Rome,mysecondaryaddress]{\textbf{A. Paiella}}
\author[Rome,mysecondaryaddress]{\textbf{F. Piacentini}}
\author[Cardiff]{\textbf{G. Pisano}}
\author[Rome,mysecondaryaddress]{\textbf{\\G. Presta}}
\author[Rome,Pasadena]{\textbf{A. Schillaci}}
\author[Cardiff]{\textbf{C. Tucker}}
\author[Milano]{\textbf{M. Zannoni}}
\cortext[mycorrespondingauthor]{Corresponding author}
\address[Rome]{Physics Department, University of Rome Sapienza, P.le Aldo Moro 5, 00185, Roma, Italy}
\address[mysecondaryaddress]{Istituto Nazionale di Fisica Nucleare, Sezione di Roma, P.le A. Moro 2, 00185 Roma, Italy}
\address[Milano]{Physics Department, University of Milano--Bicocca, Piazza della Scienza 3, 20126 Milano, Italy}
\address[Cardiff]{School of Physics and Astronomy, Cardiff University, Cardiff CF24 3YB, UK}
\address[Pasadena]{California Institute of Technology, Pasadena, CA 91125, USA}
\address[ESA]{ESA/ESRIN, Largo Galileo Galilei, 1, 00044 Frascati (RM), Italy}

\begin{abstract}

We describe the design and in--flight performance of the cryostat and the self--contained $^{3}$He refrigerator for the OLIMPO balloon--borne experiment, a spectrophotometer to measure the Sunyaev--Zel'dovich effect in clusters of galaxies. 

The $^{3}$He refrigerator provides the 0.3~K operation temperature for the four arrays of kinetic inductance detectors working in 4 bands centered at 150, 250, 350 and 460~GHz. The cryostat provides the 1.65~K base temperature for the $^{3}$He refrigerator, and cools down the reimaging optics and the filters chain at about 2~K. 

The integrated system was designed for a hold time of about 15 days, to achieve the sensitivity required by the planned OLIMPO observations, and successfully operated during the first long--duration stratospheric flight of OLIMPO in July 2018. 

The cryostat features two tanks, one for liquid nitrogen and the other one for liquid helium. The long hold time has been achieved by means of custom stiff G10 fiberglass tubes support, which ensures low thermal conductivity and remarkable structural stiffness; multi--layer superinsulation, and  a vapour cooled shield, all reducing the heat load on the liquid helium tank. 

The system was tested in the lab, with more than 15 days of unmanned operations, and then in the long duration balloon flight in the stratosphere. In both cases, the detector temperature was below 300~mK, with  thermal stability better than $\pm$ 0.5~mK.

 The system also operated successfully in the long duration stratospheric balloon flight.

\end{abstract}

\begin{keyword}
\texttt{Wet cryostat, long duration $^{3}$He sorption fridge, superinsulation, Helium vapor cooled shield, balloon-borne experiment.}
\end{keyword}
\end{frontmatter}

\section{Introduction}
Precision measurements of the Cosmic Microwave Background (CMB) require very sensitive detectors, arranged in arrays, operating at mm wavelengths. Photon--noise--limited performance \cite{Lamarre} can been achieved by cooling the optical system at cryogenic temperatures, and the detectors at sub--K temperatures, for the entire duration of the measurements.

The CMB exhibits a thermal spectrum, characterized by a single temperature, $T_{rad}$ = (2.72548 $\pm$ 0.00057)~K \cite{Fixsen}, proving that the universe underwent a dense, hot phase, as described by the standard cosmological model. This radiation is very close to be spatially isotropic. Its low level of anisotropy, as measured by many experiments in the last decades, is determined both by the fluctuations of the photon density across the Last Scattering Surface (primary anisotropies), and by the structure formation processes, leaving their imprint on the CMB along its optical path to us (secondary anisotropies). Precision measurements of CMB anisotropy at different angular scales provide invaluable information about these processes \cite{CoreDiValentino}.

Foreground emissions (galactic and extragalactic) must be minimized and removed by proper selection of the observation bands and sky regions. So, in addition to high sensitivity detectors, simultaneous measurements at different frequencies are needed. This allows to monitor and to efficiently separate different signal components, which feature different spectral shapes. 

The best way to explore the whole spectral region of interest for the CMB is to operate instruments with no hindrance from the atmospheric opacity in the microwaves, either by flying them on satellites, or by keeping them at very high altitude on sub--orbital platforms, like rockets and stratospheric balloons \cite{Masi2019}.

OLIMPO \cite{Coppo, Nati2007} is the largest telescope ever launched on a stratospheric balloon (2.6~m optical aperture). It was designed to observe the sky in the mm and sub--mm bands, using innovative Kinetic Inductance Detectors (KIDs)\cite{Paiella2018, Paiella2019, WOLTE13} and a plug--in Differential Fourier Transform Spectrometer (DFTS) \cite{Schillaci, PdB2012}. The main goal of the experiment is to measure the secondary CMB anisotropy known as Sunyaev--Zel'dovich (SZ) \cite{SZ72} effect, in the direction of selected clusters of galaxies. To this purpose, it covers four frequency bands, centered at 150, 250, 350 and 460~GHz: these frequencies match the negative, zero, and positive regimes of the SZ spectral regions.

The limited electrical power available on the payload, and the extreme temperature and pressure of the stratospheric environment, 
do not permit the use of a dry mechanical cryocooler. So we opted for a wet configuration with liquid cryogens. Moreover this configuration allows to minimize the vibrations and consequent microphonic noise in the detectors.
The chosen configuration provides the needed lifetime for the cryostat, without requiring any maintenance, nor electrical power. With careful design, it matches the forecast experiment duration at float: about 15 days for a polar stratospheric long duration balloon (LDB). 

In this paper we describe the optimization and performance of the wet cryostat and the closed cycle $^{3}$He refrigerator of OLIMPO.

Starting from the experience of BOOMERanG \cite{Masi}, the OLIMPO cryostat has been designed to have four different temperature stages: the external shell, the nitrogen tank, the helium vapor cooled shield, and the helium tank. These stages are at gradually decreasing temperatures from the outside to the innermost insert, and they are thermally--insulated by structural supports made of fiberglass.

The room--temperature external shell has been designed to sustain a very low pressure in the internal volume and to be the first shield to the radiative load on the system.

The nitrogen tank, which operates at 77~K at standard pressure, has been designed with a bell shape, and plays an essential role for the inner stages of the cryostat providing heat--sinking for the mechanical and electrical connections from the outside, and an envelope against radiation loading. 

Radiation shielding is further improved through multi--layer \emph{superinsulation}\footnote{Cryolam NRC--2, Metallized Products, Inc., Winchester, MA, USA.}.

Vapors from the $^{4}$He boiloff flow through a serpentine welded on a copper shield placed between the two tanks, cooling it at an intermediate temperature, close to 35~K (with a temperature of the nitrogen tank at 77~K) and close to about 21~K in flight  (when the temperature of the nitrogen tank is around 52~K).

The $^{4}$He stage has been built with a cylindrical shape, where the available liquid reduces to about 50~$\%$ of the full volume after the pump--down operation needed to reach a temperature of 1.65~K.
The main purpose of the $^{4}$He tank in the OLIMPO cryostat is to cool down the reimaging optics and to provide the thermal reference stage for the condensation point of the $^{3}$He evaporative sorption refrigerator.
This refrigerator has been custom--designed to work on the stratospheric balloon payload and to cool to 300~mK the focal planes of the experiment for about two weeks without being recycled.

\section{Cryostat design}

\begin{figure}[!ht]
\centering
\includegraphics[width=1\textwidth]{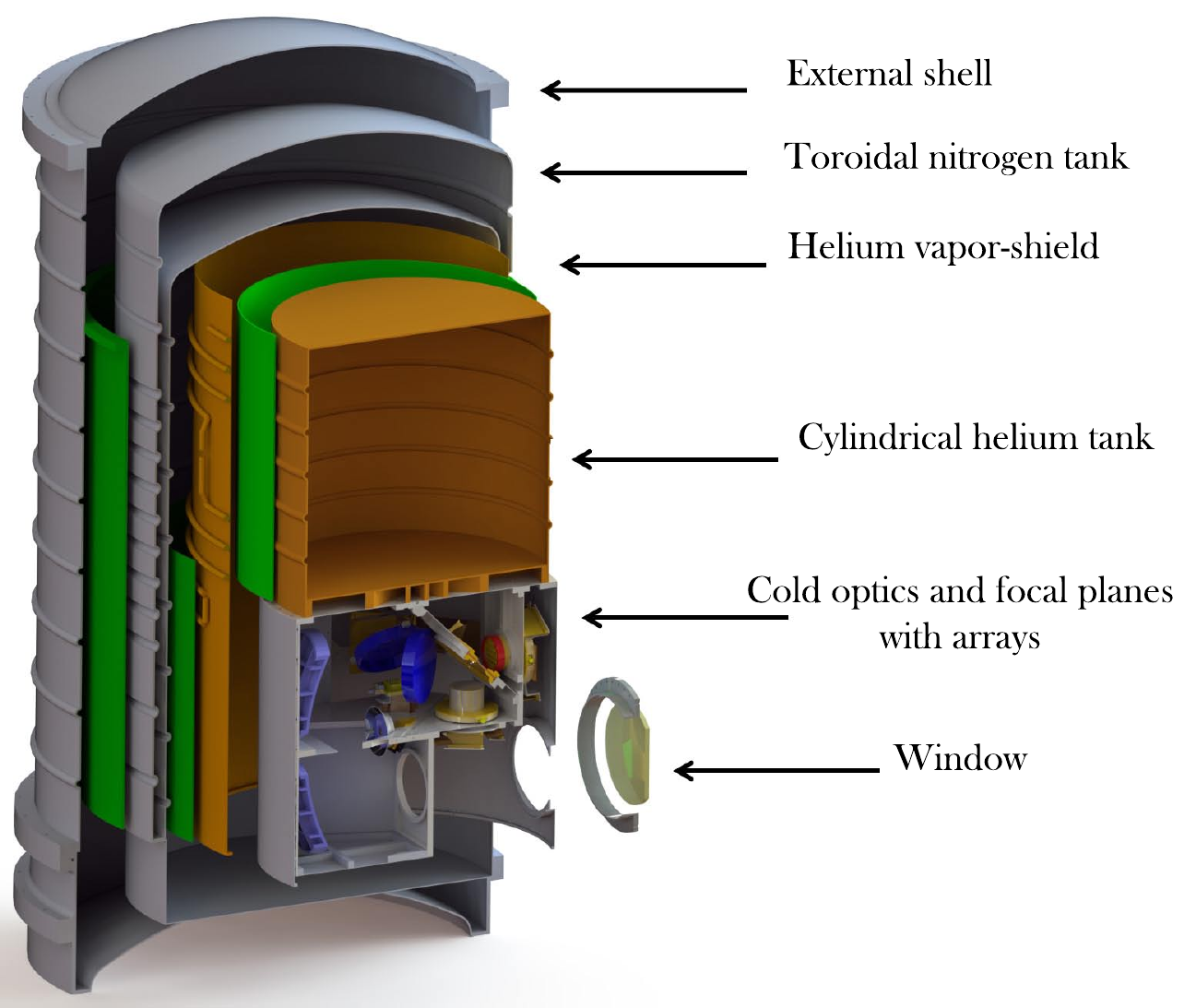}
\caption{\label{cryostat} Exploded sections of the cryostat tanks (supports removed). We can see: the external shell with the window, the toroidal nitrogen tank, the helium vapor--shield, the cylindrical helium tank, the fiberglass support in green, the \emph{cold optics} and the focal planes with the arrays. The fill and vent lines have been removed for clarity, and are discussed in \S \ref{par:ENS} and \S \ref{par:HS}. }
\end{figure}

A general drawing of the cryostat is shown in figure \ref{cryostat}.

\subsection{External shell and liquid nitrogen tank}\label{par:ENS}

The external shell and toroidal liquid nitrogen (LN2) tank, with a volume of 74~liters, are the same as those used for the cryostat \cite{Masi} of the LDB BOOMERanG experiment \cite{Crill, Masi2006}. Both the external shell and the nitrogen tank are made of 6061 aluminum to limit their mass. The external shell has been modified changing the position of the window for radiation input, since in OLIMPO the window is located on the side wall to couple the cold reimaging optics to the focal plane of the Ritchey--Chretien room temperature telescope.
The room--temperature shell is a vacuum--tight aluminum vessel able to maintain a pressure below $10^{-7}$~mbar in the internal volume. It represents the first shield against the radiative load on the receiver, thanks to the low emissivity of the inner surface. 

The LN2 tank uses the high latent heat of evaporation ($199 \frac{kJ}{kg}$ at p about 1~bar, versus $21 \frac{kJ}{kg}$ for $^4$He)
to intercept the radiation loading from room temperature, through the evaporation of a comparatively limited amount of cryogen.
Two stainless steel tubes with corrugated inserts are used as the fill and vent lines to the nitrogen tank. They have an outer diameter of 22 mm and 0.2 mm thickness, to minimize the conductive thermal loading. The corrugation segments increase the length thus reducing the heat load from the outer shell.

Differently from the BOOMERanG cryostat, where kevlar cords were used, in the OLIMPO cryostat the LN2 tank is connected to the room temperature shell and supported by a single fiberglass (G10) cylinder, 4~mm thick, 800~mm long and 674~mm in diameter, as shown in figure 1. This support provides increased stiffness, which is needed due to the telescope elevation range between $0^o$ and $55^o$, while preserving a reasonably low heat load of around 0.5~W during the flight.

Since the atmospheric pressure at stratospheric altitude is a few mbar, liquid nitrogen in the tank cools further down to about 52~K, changing from liquid to solid state and possibly loosing thermal contact with the tank walls. To cope with this effect, we added several internal copper braids, to ensure the thermal contact between the solid nitrogen and the nitrogen tank wall. The performance of this solution is briefly discussed in \S  \ref{sec:flight}.

\subsection{Superinsulation}

Multi--layer insulation (\emph{superinsulation}) is implemented between the external shell and the nitrogen tank in order to reduce the radiative load. For $n$--layers of \emph{superinsulation}, where each layer is made of highly reflective material on one side and moderately emitting material on the opposite side, the theoretical radiative power load on the innermost surface is:

\begin{equation}
\dot{Q}_{rad} = \bigg(\frac{1}{n+1}\bigg) \Big(\frac{\epsilon}{2}\Big) \sigma A \Big(T_{n+1}^{4}-T_{0}^{4}\Big)
\label{1}
\end{equation}
in the approximation $A_{i}=A$ and $\epsilon_{i}=\epsilon \ll 1$, where $A$ is the surface area, $\epsilon$ is the surface emissivity and $\sigma = 5.67 \cdot 10^{-8} Wm^{-2}K^{-4}$ is the Stefan--Boltzmann constant. 

However, in practice there is an extra conductive thermal load, due to a shallow contact between the layers, and an extra convective possible thermal load, due to the density of the layers which makes more difficult to remove the gas trapped among them. 
 
The number of layers has been chosen as a tradeoff between reduction of radiative load and available volume. We installed 30 layers of \emph{superinsulation}: 15 between the external shell and the fiberglass support tube, and 15 between the fiberglass support and the LN2 tank, for a total thickness of the superinsulation blankets of $\sim$ 15~mm; 20 layers covered the top and bottom bases of the LN2 tank, with a thickness of $\sim$ 10~mm. 

By measuring the boiling off rate of the liquid nitrogen before and after the assembly of the \emph{superinsulation}, we have observed a heat load reduction from 52~W to 8.5~W. In the flight configuration, the boiloff rate is about 0.2~l/h, which guarantees a nitrogen hold time of about 15 days for a 74~l reservoir.

\subsection{Helium vapor cooled shield (VCS)}\label{VCS}
 
\begin{figure}[!ht]
\centering
\includegraphics[width=0.9\textwidth]{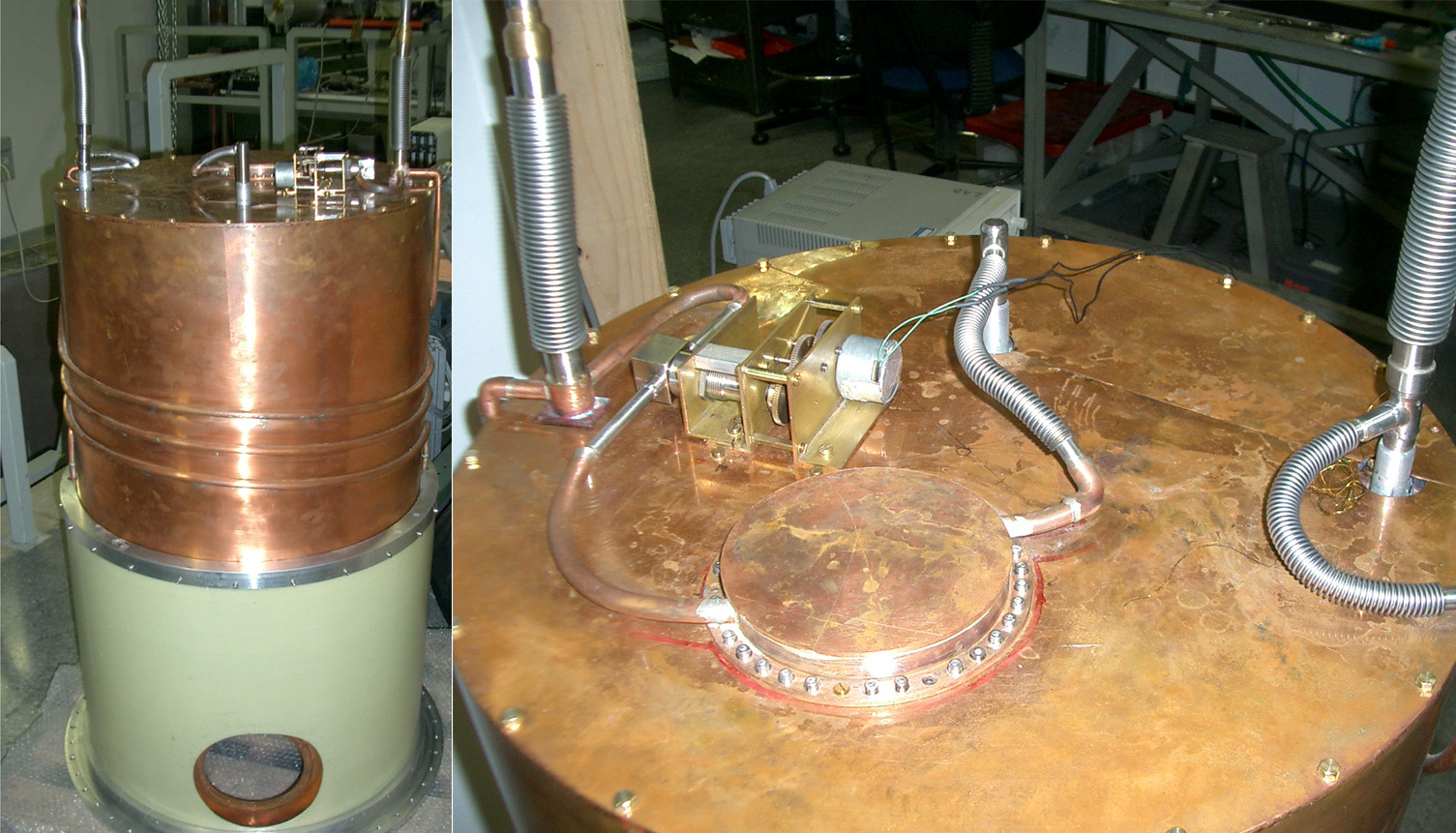}
\caption{\label{schermo_vap} Pictures: on the left the Helium vapor cooled shield, the fiberglass tube support, the serpentine and the exchanger. On the right detail of the exchanger, the valve and the mechanism used to operate the valve, regulating the flux of the vapors.  }
\end{figure}

The Nitrogen tank supports a VCS, cooled by the cold gas evaporating from the liquid $^4$He (L$^4$He) tank. The fiberglass tube supporting the VCS from the LN2 tank is 2.5~mm thick, 465~mm long, and 550~mm inner diameter (see figure \ref{schermo_vap} on the left); the fiberglass tube supporting the L$^4$He tank from the VCS is 2.5~mm thick, 440~mm long and 478~mm inner diameter.

The main purpose of this shield is to absorb radiated (mainly) and conducted heat coming from the nitrogen tank, reducing the total heat load on the L$^4$He tank. The VCS is a self regulating system: if the shield is too warm, the evaporation rate of helium in the L$^4$He tank increases, due to higher loading from the shield, thus producing a larger flux of cold vapors and therefore providing additional cooling for the shield itself. 

This shield is split in two copper cylinders, interrupted by a center flange where both fiberglass cylinders are bolted on. A copper serpentine, collecting the flow of cold gas evaporated from the L$^4$He tank, is hard soldered around the cylinders. The length of the serpentine is $\sim$ 11~m and the internal diameter is 8~mm.  

Since its high impedance might produce pressure oscillations during the liquid helium pump--down process, a lower impedance heat exchanger was assembled in parallel to the serpentine, with a  motorized valve regulating the fraction of $^{4}$He vapors flux flowing along the two paths. The shield and the valve with the heat exchanger are visible in figure \ref{schermo_vap}.

The valve is a \emph{Swagelok} model SS--6BW--TW\footnote{https://www.swagelok.com}, made of stainless steel. The valve was degreased with ultrasonic cleaning to prevent it from getting stuck when cooled at cryogenic temperatures. The valve is rotated by a \emph{GlobeMotors} DC motor with bronze bushings, which have been rebored ($+100$~$\mu$m on the diameter) for cryogenic temperature operation. In \S \ref{par:labo} we shortly report on the performance of this system.

\subsection{Liquid Helium Tank}\label{par:HS}

The L$^{4}$He tank cap volume is 64~liters. The boiling temperature of helium drops from 4.2~K at standard pressure to below 1.65~K when the pressure of the vapors is reduced to a few mbar in the stratosphere. 

The innermost and coldest stages of the OLIMPO cryostat (including the refrigerator, the detectors, the cold optics and the optical filter chain) have been thermally and mechanically designed to take advantage of the stratospheric environment, and thus cannot be properly operated when the pressure over the helium reservoir is higher than a few mbar.
Therefore the L$^{4}$He reservoir must be pumped in the laboratory to operate the $^3$He refrigerator, allowing for pre--flight detectors validation and calibration. Just before launch, the mechanical rotary pump used for maintaining the low pressure on the L$^4$He bath is disconnected, after having shut the vent line by means of a remotely operated valve. In this way the pressure and temperature of the L$^4$He bath drift slightly up while, in a few hours, the payload reaches the floating altitude. There, the vent valve is re--opened, and the low ($\sim 300$~Pa) outer pressure maintains the liquid at a steady temperature of 1.65~K. This procedure avoids a large loss of cryogen which would be unavoidable if pumping the He bath in uncontrolled conditions during the ascent. The efficiency of this procedure has been confirmed in flight, as described in \S \ref{sec:flight}. 

Even after an optimized pump--down (about 1.5~mbar/min) and tuning of the pumping line impedance, $\sim 50~\%$ of the L$^{4}$He is used to cooldown the remaining $\sim 50~\%$ and the attached equipment. The final temperature of the L$^{4}$He bath is below 1.65~K, enough to efficiently condensate the $^{3}$He in the refrigerator and to cool the optics box and the coldest stages of the optical filtering chain.

In order to interface the bottom of the helium tank directly to the cold optics we decided to make both the top and bottom flanges flat. The thickness \emph{t} of the tank was chosen according to the following formula \cite{Conte} :

\begin{equation}\label{thick}
t = 0.866 \cdot r \cdot \sqrt[]{\frac{P \cdot X}{R }}
\end{equation}

where $r$ is the radius of the tank, $P$ is the pressure applies from the outer side, $X$, the safety factor, and $R$, the breaking strength.

For a standard pressure of $\sim10^{5}$~Pa, radius $r$ = 0.225~m, $X$ = 5 and $R \sim 3\times10^{8}$~Pa for the aluminum we obtain a thickness of about 8~mm. For the fabrication of the flange we started from a 30~mm thick flat and lightened it by milling out most of the material, leaving a network of thick ribs, defining many areas with typical side of 100 mm, and reinforcing a 5 mm thick flat skin.

We decided to add some ribs around the tank and some ribs on the internal side of the bottom flange of the tank, to optimize the thermal contact between the liquid and the fridge.

\begin{figure}[!ht]
\centering
\includegraphics[width=0.8\textwidth]{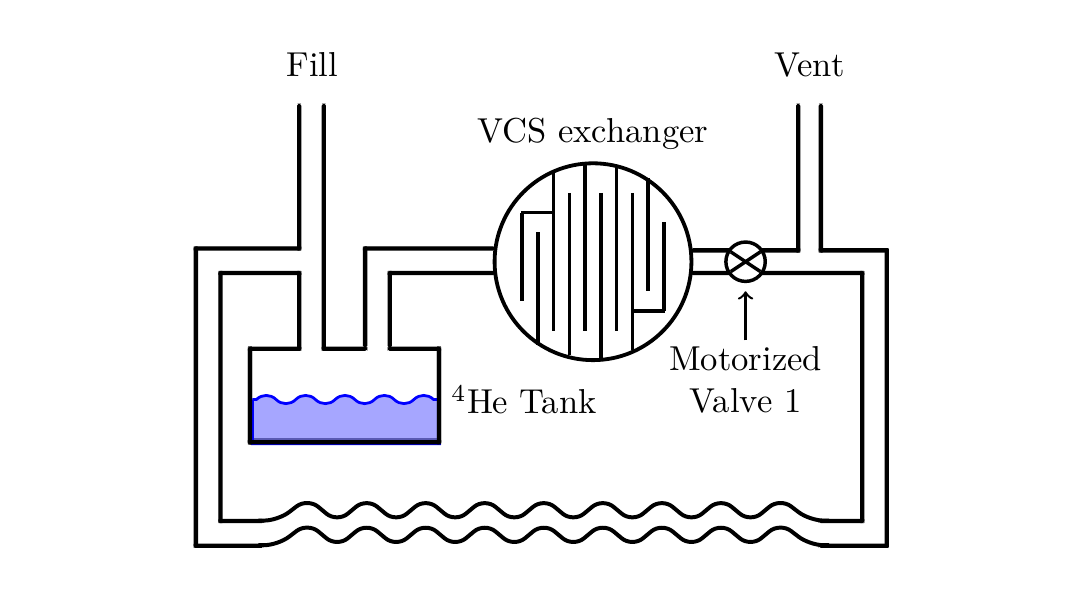}
\caption{\label{Exchanger} Schematic diagram of L$^4$He tank connections. Two steel tubes connect the tank to the exterior: one for Fill, one for Vent. The former is directly welded to the helium tank, while the latter is split in two parallel paths. One is a serpentine hard soldered on the VCS, and one is a bypass heat exchanger on the top cover of the VCS. The flow through the heat exchanger can be regulated by means of motorized valve 1.
}
\end{figure}

To fill with liquid Helium and to provide a vent for the vapors, we use two tubes, made of stainless steel with a thickness of 0.2~mm and a diameter of 12.7~mm. 
The evaporation tube is split into two directions, one towards the serpentine and the other towards the valve, then towards the exchanger flowing into the Helium tank.
The filling tube is connected straight to the tank, and is also connected to one of the two serpentine outputs, as shown in fig. \ref{Exchanger}.
The fill and vent tubes are interrupted with a 200~mm long section where 185~mm $\emptyset$ stainless steel bellows is inserted. This reduces the heat load and allows for small shifts of the two ends, due to thermal contraction effects.

The geometry of the system and the heat loads on the helium bath, mainly from cables harness and fiberglass tube support,
are such that the equilibrium of the VCS is expected to be reached at a temperature of 35~K (with the LN tank at 77~K).
The fiberglass tube connected between the VCS and L$^{4}$He is 2.5~mm thick, 45~mm long and 476~mm in diameter.

\subsubsection{External Plumbing}

\begin{figure}[!h]
\centering
\includegraphics[width=0.8\textwidth]{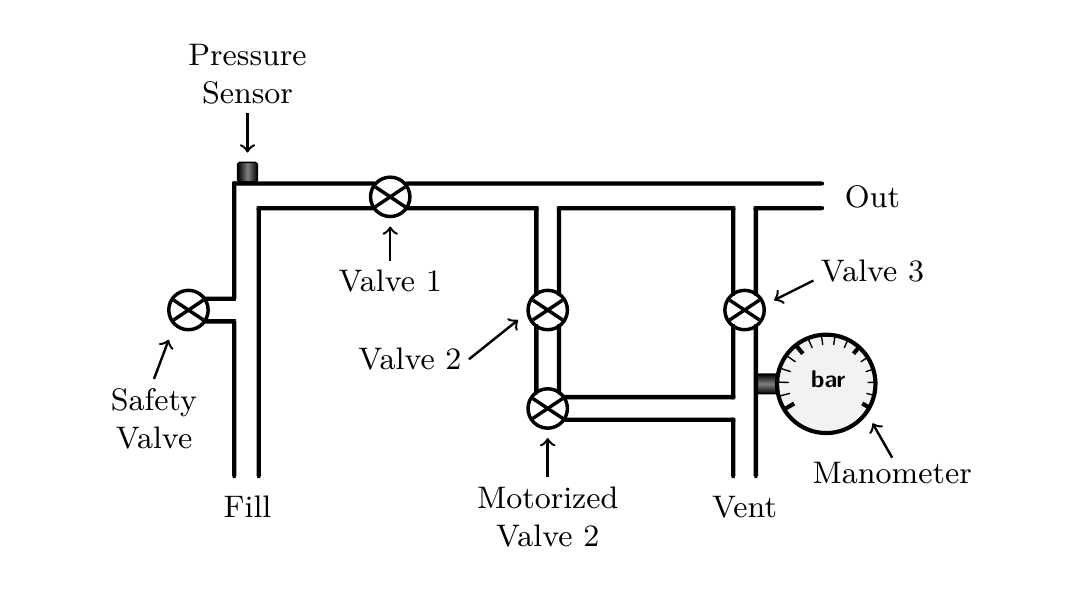}
\caption{\label{Plumbing} System design of external Plumbing for helium tank. The vapors flow through the vent line while the fill line is closed. When the instrument is in pre--flight configuration and during ascent the valves 1, 3 and motorized valve 2 are closed, while valve 2 is open. When the system is at float in the stratosphere, the motorized valve 2 is opened and pumping is restored, due to the external pressure of about 3~mbar. Valve 2 is used only in the laboratory, during pump--down (see text).}
\end{figure}

The filling and venting tubes for the helium tank are externally connected through a plug--in network of valves and tubes, which allows to adjust on--the--fly the line impedance of the pumping system used to  lower the pressure over the liquid (see figure \ref{Plumbing}). This system also features a remotely--operated motorized valve, used to seal the tank before launch and open it to the outside pressure when the cryostat is at float in the stratosphere. The motorized valve is a critical subsystem, and potentially a single--point failure. For this reason we avoided unnecessary complexity in the motor control electronics, which allows only for two states of the valve: fully open or completely shut, without any control of the intermediate positions. However, in the laboratory, during the L$^4$He pump--down process, we have to fine--tune dynamically the pumping speed, to avoid Taconis oscillations. For this reason we have added a manually operated valve (valve 2 in figure \ref{Plumbing}) in series to the motorized valve. The manually operated valve 2 is used only during the pump--down and left wide--open when the pump--down is completed.

\section{Window and Filters}

The OLIMPO focal plane features a large throughput ($\sim$2~cm$^{2}$sr). This means that the stack of filters must limit out--of--band radiation which would represent a very large heat load.
The radiation enters the cryostat through the window and encounters the first thermal shader, located right behind it. The window is 8~mm thick and it is made of HDPE (High Density Polyethylene), a material with high transmittance and low reflectance at mm wavelengths \cite{D'Alessandro}, thus ensuring minimum excess background on the detectors, while preserving excellent mechanical strength and vacuum tightness. 
In order to avoid interference due to internal reflections, an anti--reflection coating made of porous Teflon on a low--density polyethylene substrate was applied on the HDPE surfaces. The 130~mm aperture of the window is determined by the configuration of the cold optics and by the size of the radiation beam.

\begin{figure}[!ht]
\centering
\includegraphics[width=1\textwidth]{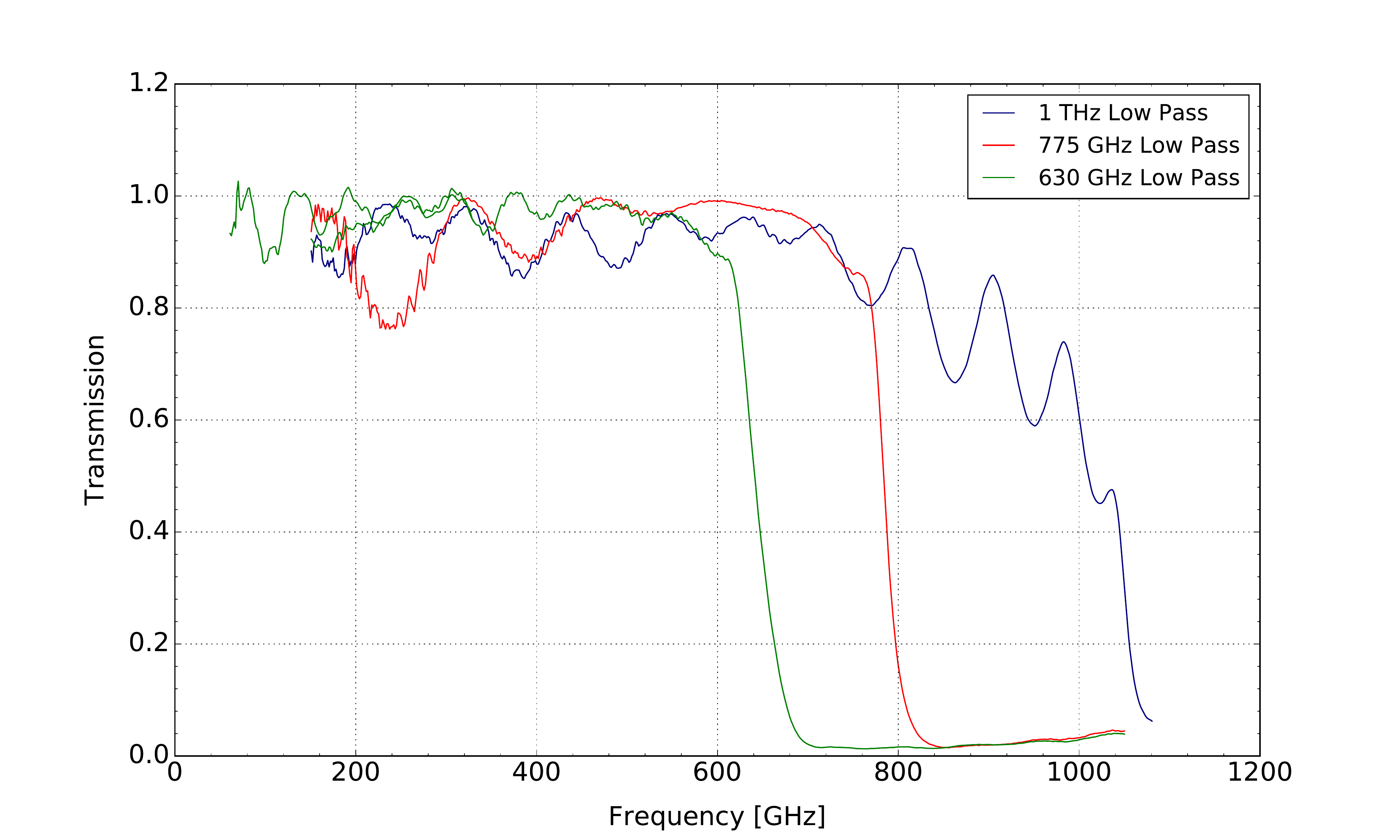}
\caption{\label{Filtri} Transmission spectra of the low--pass filters mounted in the OLIMPO cryostat.}
\end{figure}

Then, the radiation passes through the second thermal shader, the 1~THz low--pass filters, and the third thermal shader, all three anchored to the LN stage. After passing through the fourth thermal shader and the 775~GHz low--pass filter mounted on the vapor $^{4}$He shield at 35~K (in--flight the temperature is about 21~K), the incoming radiation encounters the 630~GHz low--pass filter mounted on the \emph{optics box} shield at about 2~K. Fig. \ref{Filtri} shows the transmission spectra of the three low--pass filters mounted in the OLIMPO cryostat.

\section{Laboratory performance of the cryostat}\label{par:labo}

The cryostat has been assembled and tested in the lab in more than 30 complete cooling cycles. A 40~m$^3$/h rotary pump has been used to pump on the L$^4$He bath, reaching a steady state temperature, after a $\sim$12 hours long pumpdown, of 1.65~K. 

We have experimented different settings for the gas flow in the long VCS serpentine and in the top low--impedance heat exchanger described in section \S \ref{VCS}. The best configuration (no oscillations, temperature gradient of a few K across the shield) was found with the valve in a midway position, allowing for a non--negligible gas flux in the long VCS serpentine, while most of the gas ($\sim$ 70~\%) flows through the top low--impedance heat exchanger.

The total heat load on the L$^{4}$He in steady--state conditions (and with LN at 77~K) is of the order of 80~mW, as estimated from the measured L$^4$He evaporation rate (0.10~l/h). The resulting hold--time for both L$^4$He and LN is about 15 days, as confirmed by several complete operation cycles in laboratory. 

\begin{figure}
    \centering
    \includegraphics[width=1\textwidth]{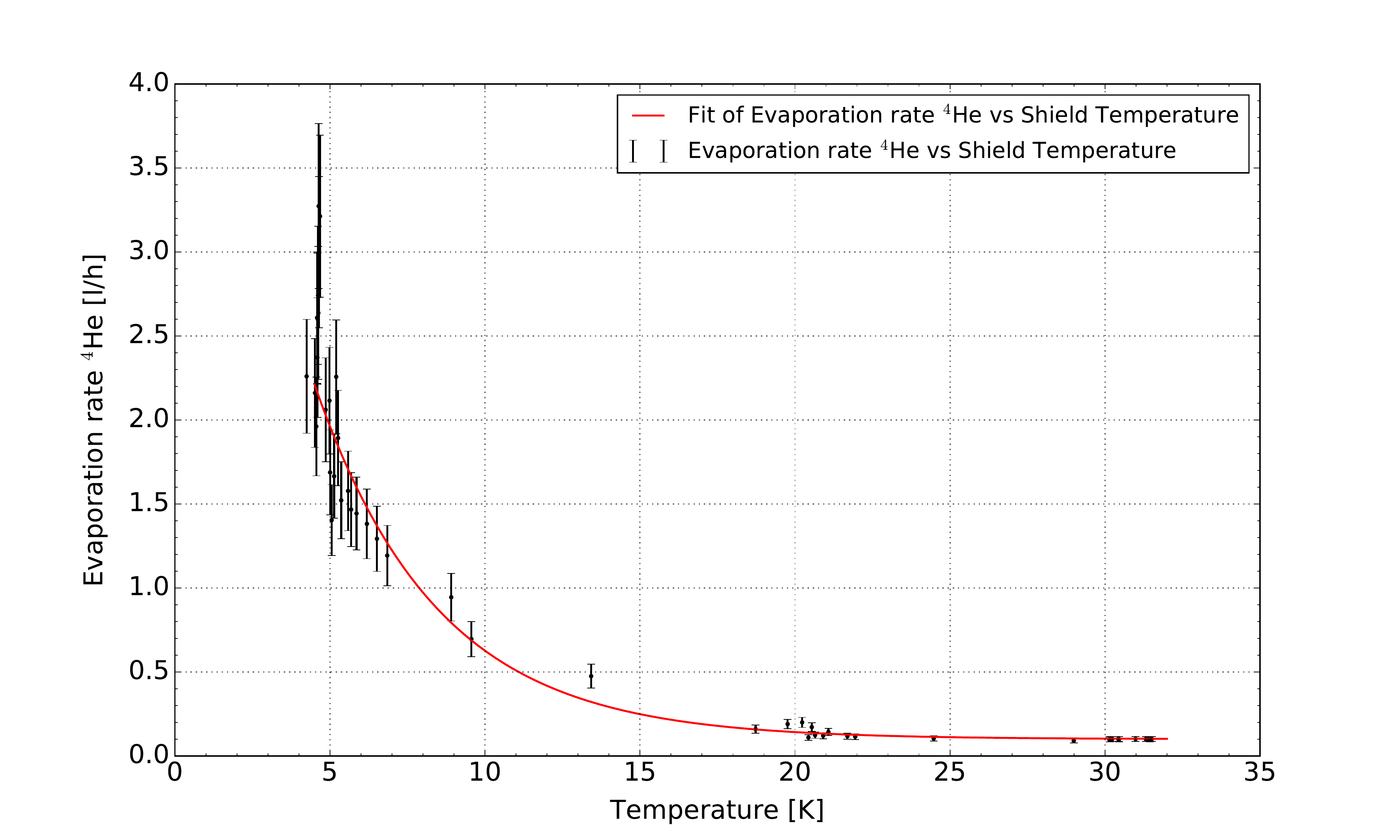}
    \caption{Measured He evaporation rate (black dots) and empirical fit (red solid line) versus vapor cooled shield temperature.
    }
    \label{Shield}
\end{figure}

The best fit of the data results with an empirical relation ($y[\rm l/h]=6.58\cdot e^{-\frac{x[\rm K]}{3.96}}+0.1$) between the shield temperature and the L$^{4}$He evaporation rate is shown in fig.\ref{Shield}, derived from measurements taken during the pump--down process (with LN$_2$ at 77~K). This relation is useful to approximately estimate the evaporation rate during the operations of the cryostat at balloon altitude, in the absence of a flux meter, since the VCS temperature is monitored continuously during the flight (see discussion in \S \ref{sec:flight}).

\section{$^{3}$He Refrigerator}

We have designed a simple single--shot $^{3}$He refrigerator, composed of a cryopump, a heat exchanger in thermal contact with the $^{4}$He tank, and an evaporator. The fridge is very similar to the one used in the BOOMERanG experiment \cite{Masi1998}, with the exception of the gas--gap heat switch, used this time to connect  the cryopump to the $^{4}$He bath when needed. The fridge contains about 40~$l_{STP}$ of $^{3}$He, required to keep the four detector arrays in their aluminum holders and their horn arrays at 300~mK for about 16~days. The total geometrical volume available inside the fridge is 1000~cm$^{3}$, but the available volume for the gas is about 80~$\%$ of that, after accounting for the volume of the carbon grains inside the cryopump. The internal gas pressure at room temperature is thus of the order of 
4~MPa, so all the parts of the evaporator must be designed to withstand high internal pressure with a reasonable safety factor.

The spherical cryopump is composed of two welded stainless steel half--spheres, with a diameter of 120~mm, and is enclosed in a  gold--plated copper shield, to prevent radiation to warm the optics box when the active carbon is heated for fridge cycling.
The thickness of the hemispheres ($T$ = 2~mm) has been calculated from the equation \cite{Conte}:

\begin{equation}
T = \frac{PDX}{5R\Psi}
\end{equation}

where $P$ is the pressure, $D$ is the diameter, $X$ is the safety factor (which is about 5), $R$ is the breaking strength of material, $\Psi$ is a form factor (23/22 for a sphere).

\begin{figure}[!ht]
\centering
\includegraphics[width=1\textwidth]{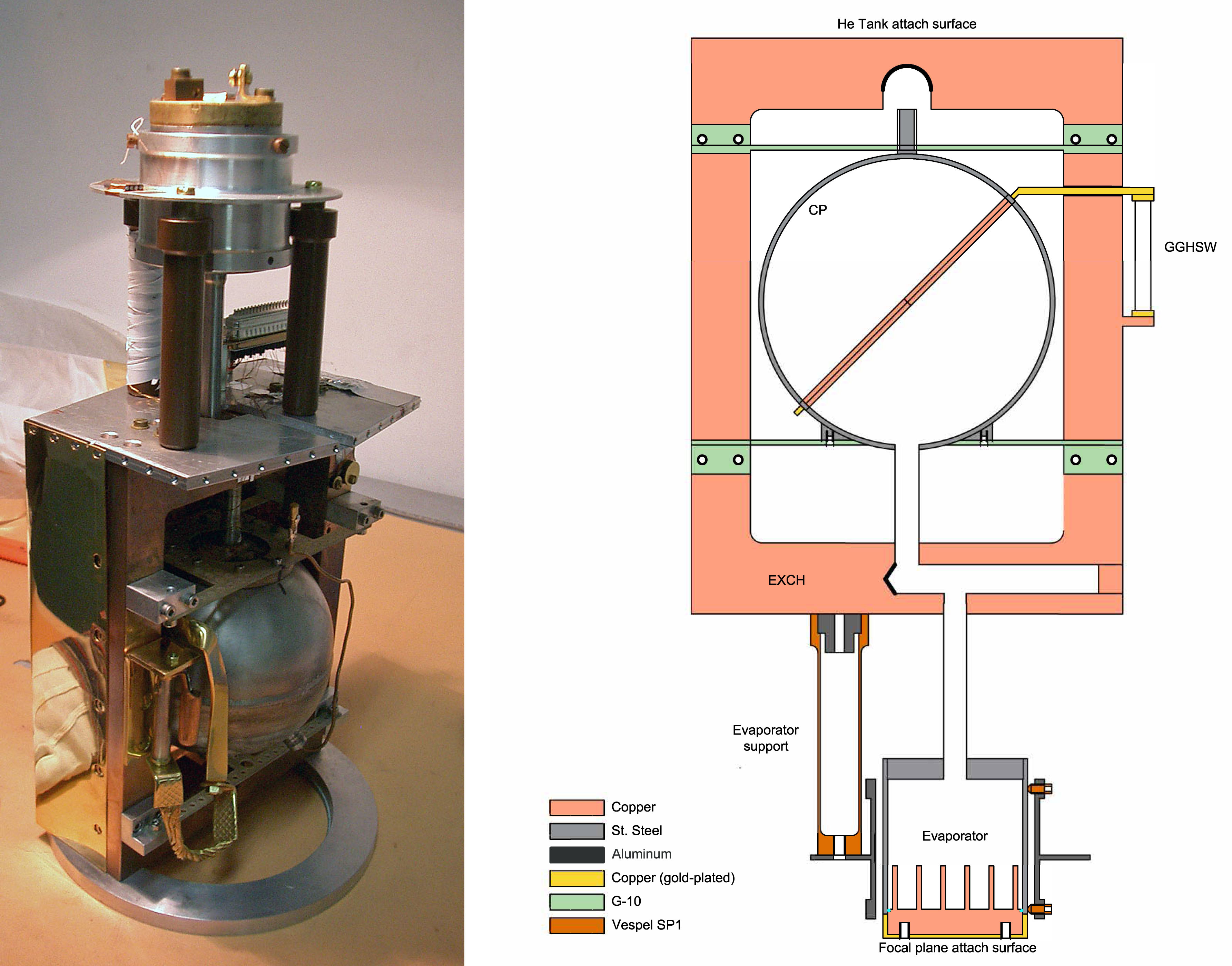}
\caption{\label{Cryopump} \emph{Left panel}: picture of $^3$He refrigerator (upside down in this picture). Note in the bottom part of the picture the spherical cryopump, with the copper link that connected the inner copper tube of cryopump with heater and heat switch, and the two fiberglass supports; in the middle the thin wall stainless steel tube connecting the cryopump to the heat exchanger at L$^4$He temperature, in the top part the aluminum structure of the evaporator, which is supported by three Vespel tubes.
\emph{Right panel}: drawing of $^3$He refrigerator. GGHSW is the gas--gap heat switch , CP is the cryopump and EXCH is the heat exchanger.}
\end{figure}

In order to increase the temperature uniformity of the carbon grains, we have inserted a copper finger, crossing the cryopump along a diameter. One end of this tube is connected to the cryopump heater resistor, and to the cryopump--exchanger thermal link, operated through the gas--gap heat switch.

Two fiberglass supports, above and below the cryopump, connect the cryopump to the copper support of the exchanger, taking care of protection against radial accelerations, as shown in figure \ref{Cryopump}. 
Further protection against radial accelerations is provided for the evaporator chamber, which is secured coaxially with the main axis of the cryostat by three insulating radial spikes (in vespel, with a heat load lower than $\sim$ 1~$\mu$W) supported by an aluminum ring structure anchored on three Vespel tubes. Moreover, thanks to the optical scheme and the consequent arrangement of the 4 detector arrays, the thermal link structure connecting the evaporator to the detector arrays ensures additional structural stiffness along three axes.

The cryopump and the evaporator are connected to the condensation section by means of thin--wall (t $\sim$ 150~$\mu m$) stainless steel tubes (9.5~mm in diameter for the evaporator--condenser section, 10~mm in diameter for the condenser--cryopump section), which prevent damage from axial accelerations, while ensuring good thermal insulation from the heat exchanger.

The safety factor to withstand the internal pressure is $X$ $\sim$ 5.
The evaporator is a 60~mm diameter stainless steel cylinder ($D$), with a thickness ($T$) of 2~mm, as computed from the equation \cite{Conte}:

\begin{equation}
T = \frac{PDX}{2R\alpha}
\end{equation}
where $\alpha$ is a coefficient depending on the quality of weldings, in our case is $\alpha \sim$ 0.8, $P$ is the pressure and $R$ is the breaking strength of material.

\begin{figure}[!ht]
\centering
\includegraphics[width=0.9\textwidth]{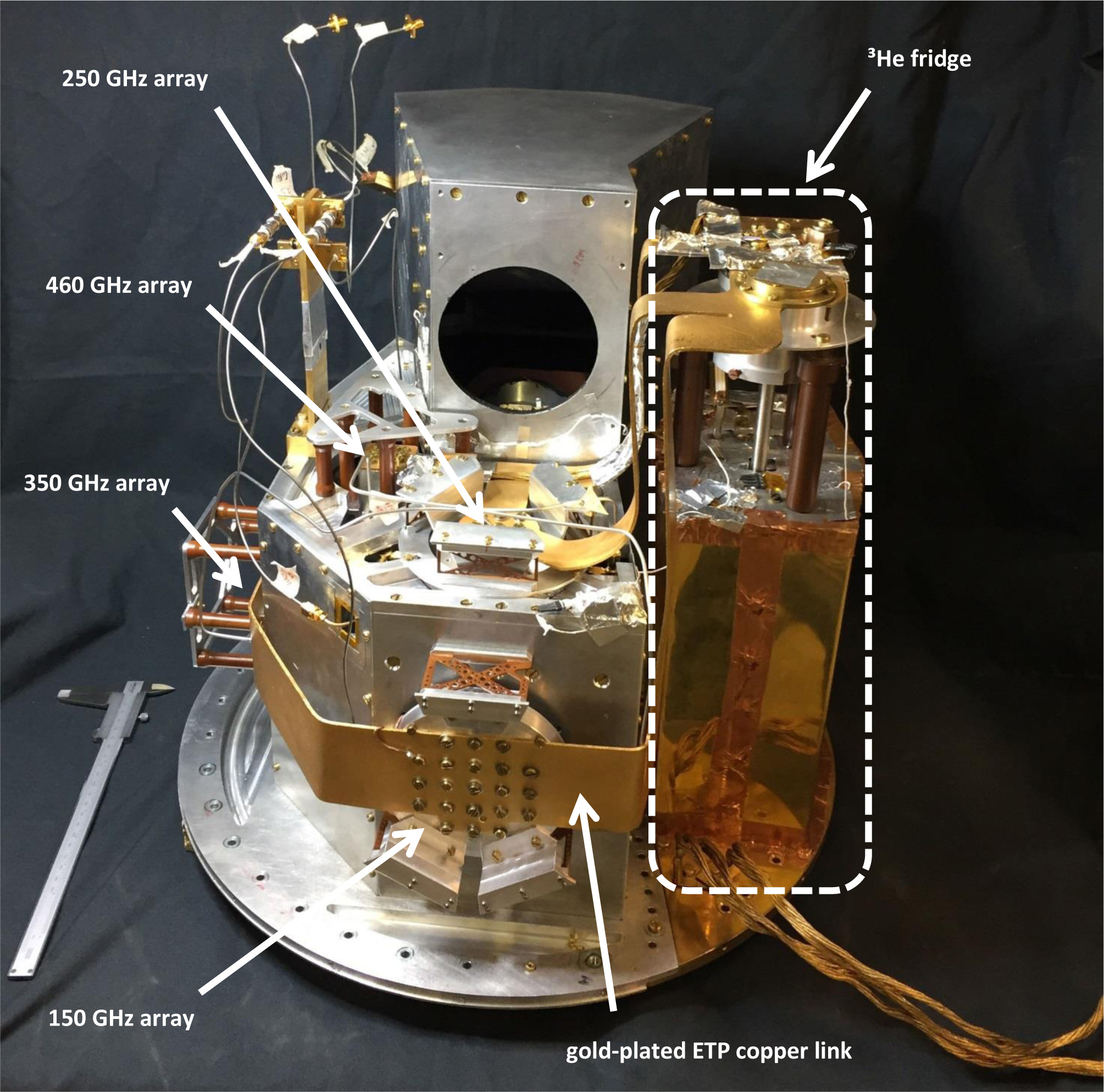}
\caption{\label{ColdOptics} Picture of the cryogenic instrument insert (shown upside down with respect to its positioning inside the L$^4$He cryostat). We can see the four arrays of detectors, the thermal link, and coaxial cables between arrays and stage at 2~K. On the right we can see the Refrigerator with its cryopump shield and its evaporator with the aluminum structural support.  
}
\end{figure}

While the thickness of the copper baseplate, 10~mm, has been computed from equation \ref{thick}, in this case the safety factor is $X$ $\sim$ 4 and $R$ is smaller than the breaking strength of copper due to welding.

The entire cryogenic system was installed at the telescope and operates in a wide range of inclinations from vertical, due to the elevation range of the telescope boresight, ranging from $e=0^o$ for calibration to $e=55^o$ at the maximum useful elevation. This means that the symmetry axis of the cryostat will change from vertical to 55$^o$ offset from vertical. These movements can change the exchange surface between the L$^3$He and the bottom flange of the evaporator, thus changing the  temperature gradient between the liquid and the detectors. To minimize this effect, the inner surface of the copper baseplate has been increased by adding a walled structure, shaped to maintain a more uniform distribution of the liquid even when tilted from vertical.

The thermal contact between the copper flange of the evaporator and the array holders is obtained by means of a gold--plated electrolytic tough pitch (ETP) copper link.

In figure \ref{ColdOptics} we can see a picture of the Cold Optics with the $^{3}$He refrigerator, the four arrays of detectors, supported by Vespel tubes for the High Frequency arrays, and by thin perforated--layers of Vespel for the Low Frequency arrays, and the thermal link connecting the refrigerator to the arrays.

\begin{figure}[!ht]
\centering
\includegraphics[width=1\textwidth]{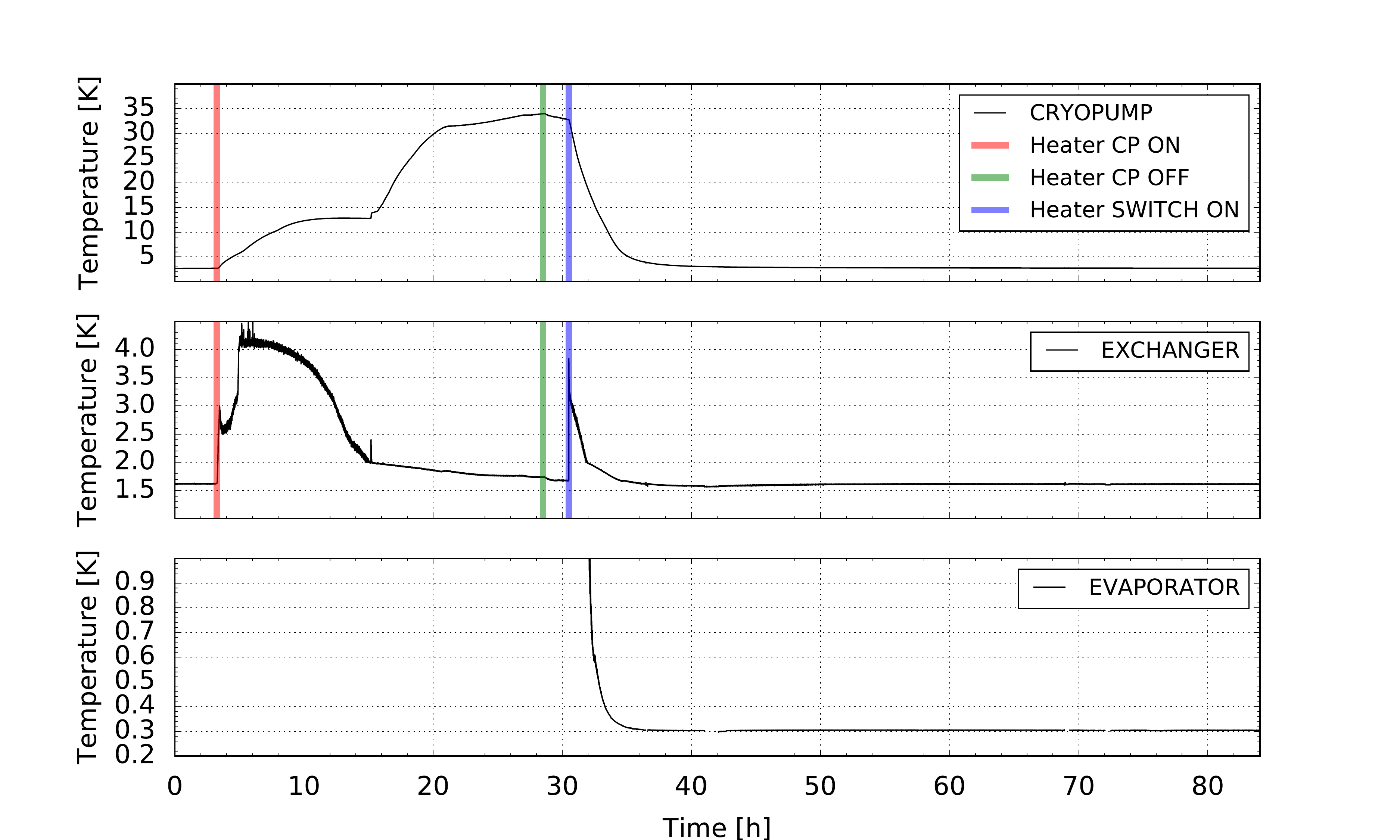}
\caption{\label{coolingcycle}Temperature of the cryopump, exchanger and evaporator during a cooling cycle.
}
\end{figure}

In figure \ref{coolingcycle} we plot the temperatures of the cryopump, heat exhanger and evaporator during a typical cooling cycle.
The cycle can start when the condensation point has been cooled by the main $^4$He bath below 1.7~K, 
the cryopump is also cold, and all the gas is adsorbed. Then electrical power is applied to the cryopump heater. The cryopump heater power is not constant, it is adjusted dynamically to optimize the desorption of the gas, while keeping the condensation point reasonably cold. In the initial phase a significant power $\sim 0.2$~W is applied to the cryopump heater, to warm it up and initiate desorption. This results in a significant temperature increase of the condensation point. After this initial phase, when most of the gas is desorbed, the heating power can be gradually reduced, while the cryopump continues to warm--up, so that the condensation point cools down. This improves the efficiency of gas condensation. The cryopump is maintained at a temperature higher than 30~K, until the temperature of the condensation point returns to 1.65~K, which ensures the condensation of the full charge of $^3$He. Finally (30.5~h in the plot), the heater is switched off, and the heat switch between the cryopump and the main $^4$He bath is closed, thus activating the cryosorption and reducing the pressure on the $^3$He bath, reaching the asymptotic temperature.

\subsection{Thermometry}

We equipped the cryostat with several cryogenic thermometers, both silicon diodes and germanium resistance temperature sensors, (as specified in table \ref{tab1}) which are needed for cryo--operation monitoring and to exclude or decorrelate systematic effects in the measurements of the CMB, due to temperature drifts.

\begin{table}[ht]
\begin{center}
\begin{tabular}{c|c|c}
\hline \hline
Stages & Diode & germanium \\
\hline \hline
77~K & 1 & 0 \\

35~K & 2 & 0 \\

2~K & 2 & 0 \\

Refrigerator & 3 & 2 \\
\hline \hline

\end{tabular}
\caption{Type and number of thermometers of the OLIMPO cryostat, mounted at different stages.}
\label{tab1}
\end{center}
\end{table}

The silicon diodes used into the cryostat are DT--670 provided by \emph{Lake Shore Cryotronics, Inc.}\footnote{https://www.lakeshore.com}. For the $^3$He refrigerator temperatures, we used germanium resistance temperature sensors, also from \emph{Lake Shore Cryotronics, Inc}, model 

GR--200A--1500, for the range of temperatures 6$\div$1.4~K and GR--200A--100 for the range of temperatures 2.1$\div$0.259~K. These are biased with an ultrastable 1~nA sinusoidal current (AC at 3.33~Hz) and readout by means of synchronous demodulation of the voltage signal across the thermistor. Custom readout boards provide temperature measurements with absolute accuracy of $\pm$0.5~mK at 300~mK and $\pm$0.1~K at 1~K for the germanium resistance temperature sensors, and $\pm$0.5~K in the 4.2$\div$77~K range for the silicon diodes. The performance of the KID detectors depends strongly on their operating temperature and the temperature of the radiative environment. The accuracy of the thermometry in flight is thus necessary to confirm that the detectors operate in conditions within the range of temperatures where they have have been calibrated in the laboratory.

\subsection{Refrigerator load curve}

\begin{figure}[!ht]
\centering
\includegraphics[width=1\textwidth]{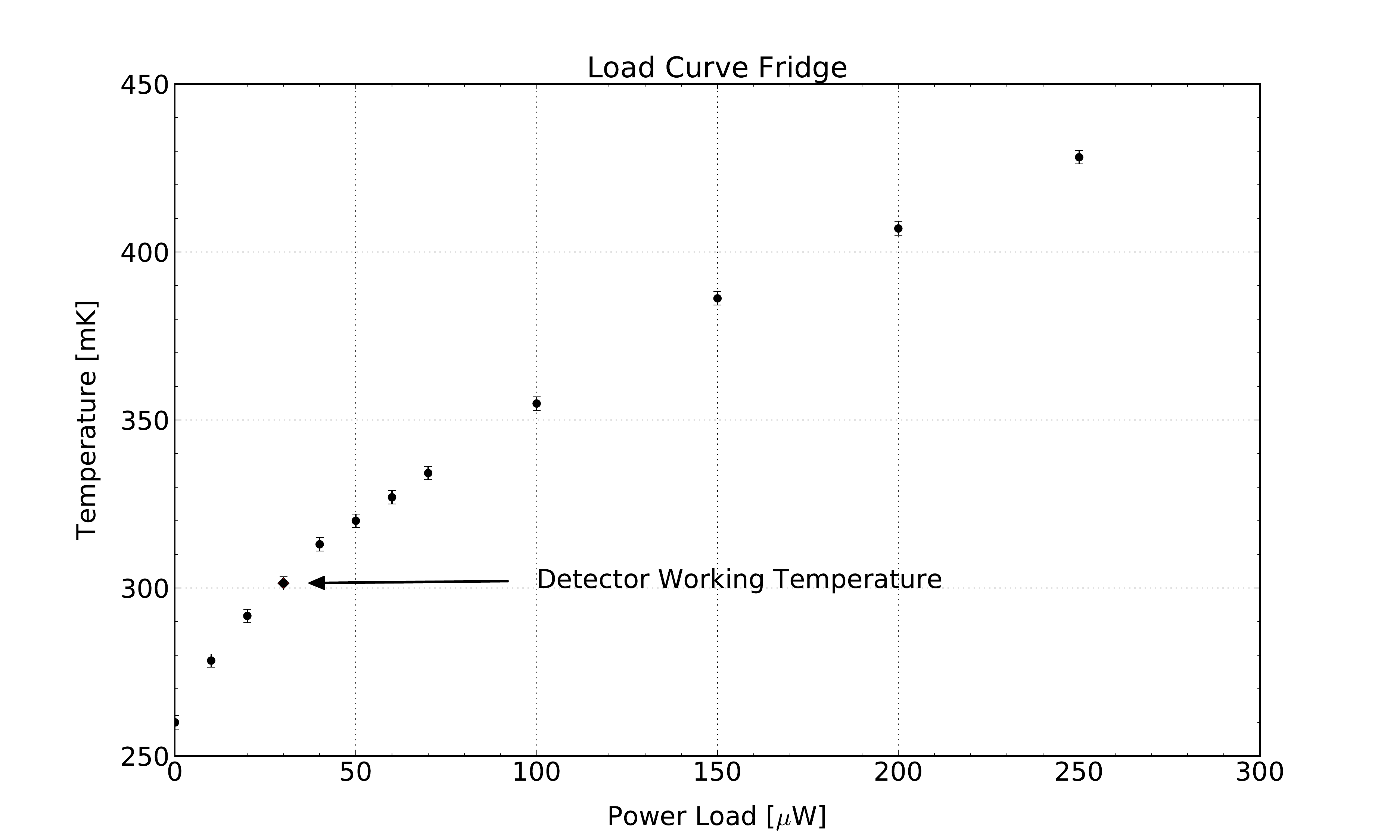}
\caption{\label{LoadCurve} Load Curve of the OLIMPO refrigerator: evaporator temperature vs external heat load power. With the nominal heat load of 30~$\mu$W the temperature of the evaporator is close to 300~mK.}
\end{figure}

In order to characterize the performance of the refrigerator we performed a specific cooling cycle, without the four arrays and their cables (in terms of thermal load, without 30~$\mu$W). We mounted a stable resistor (100~k$\Omega$) on the evaporator to dissipate controlled power levels, and measured the equilibrium temperature for different loads. 
For the measurement of the load curve, see figure \ref{LoadCurve}, the maximum temperature variation of the exchanger was about 10~$\%$ of its base temperature (1.65~K).

\section{In--flight performance}\label{sec:flight}

At 05:57 GMT on July $14^{th}$ 2018, after regular fill and pump--down procedures, the motorized valve of the $^{4}$He tank was closed by command in preparation of the launch of the experiment.
The vapor pressure of the $^{4}$He bath started to increase slowly, and, as a consequence, the temperatures of the $^{4}$He bath and of the $^{3}$He evaporator increased up to 1.71~K and 307~mK, respectively.

At 07:07 GMT on July $14^{th}$ the OLIMPO payload was launched from Longyearbyen (Svalbard, Norway. 78°13′~N~15°33′~E).

\begin{figure}[!ht]
\centering
\includegraphics[width=1\textwidth]{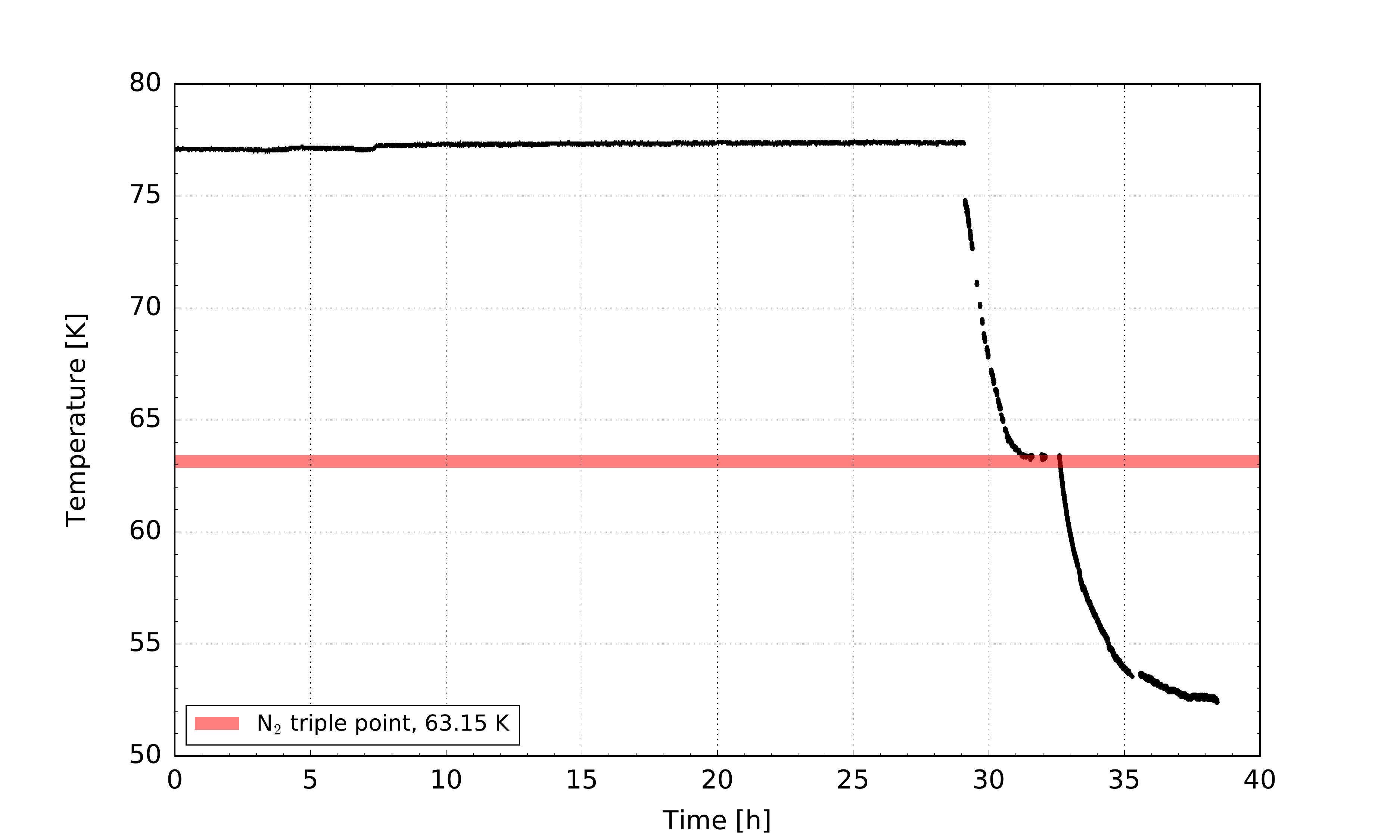}
\caption{\label{T_N2}  Temperature of LN tank before launch (first 29 hours, with the liquid evaporating at standard pressure), during ascent, and at the beginning of floatation at 37.8~km (subsequent hours). The passage through the triple point of nitrogen is evident at hour 32.5.}
\end{figure}

In figure \ref{T_N2} we plot the liquid nitrogen temperature from before the launch until floatation, where the temperature stabilized at $\sim$ 52~K, and nitrogen is solid. This low temperature of the LN stage is beneficial for the heat load of the LHe stage and for the duration of the mission. Note that the temperature of the nitrogen stage during the flight is measured to be less than 2~K above the sublimation temperature of solid nitrogen at the floating pressure. Since the heat load on the nitrogen tank is estimated to be $\sim$ 8~W, the effective heat conductance of the Cu braids in the LN tank is $> 4$~W/K.

At 11:37 GMT, at an altitude of about 35~km, the temperatures of the $^{4}$He bath and of the $^{3}$He evaporator were 1.90~K and 320~mK respectively, and the motorized valve was opened by telecommand. In this configuration the $^{4}$He tank was connected directly to the stratospheric pressure (about 3~mbar) and the $^{4}$He bath vapour pressure and its temperature started to decrease.

\begin{figure}[!ht]
\centering
\includegraphics[width=1\textwidth]{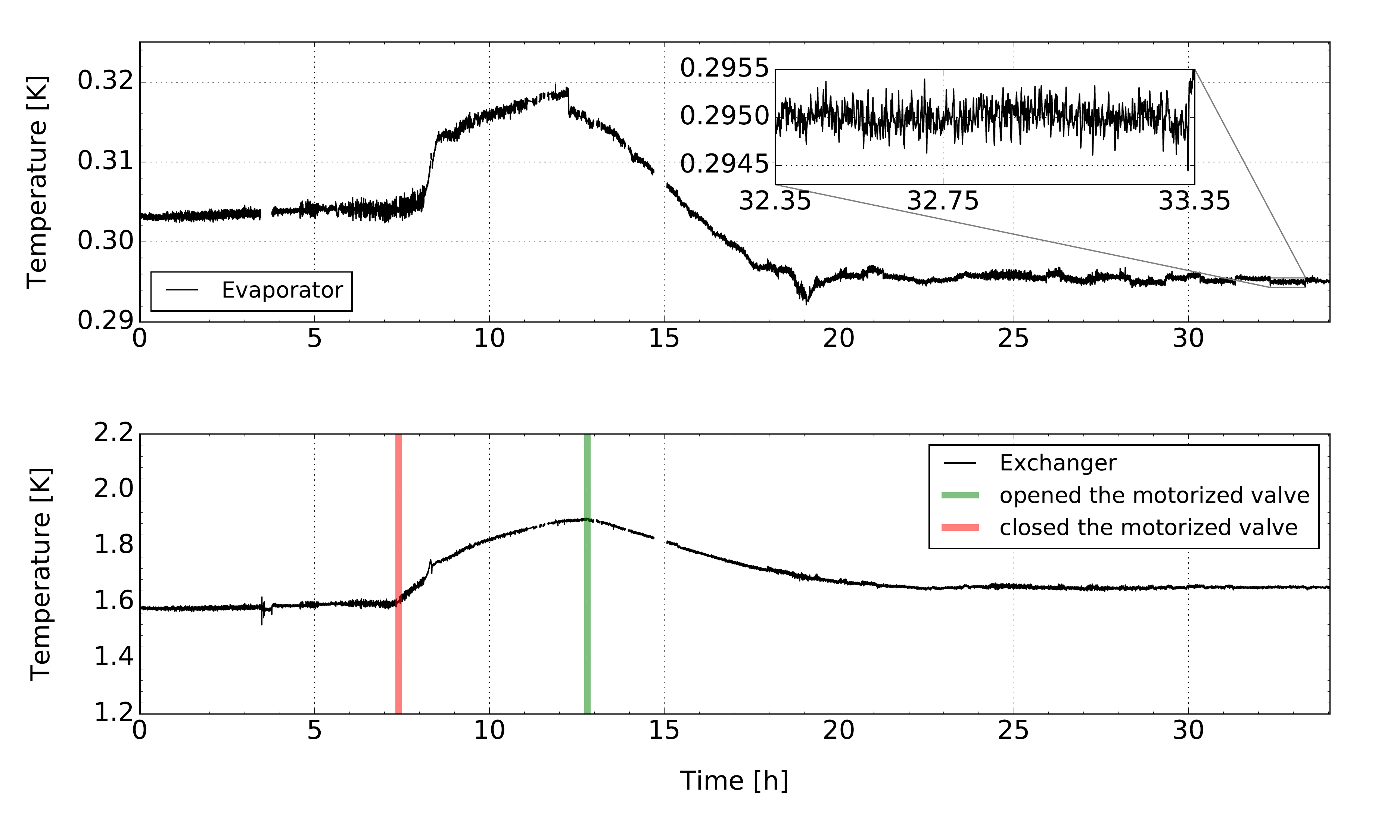}
\caption{\label{Temp_Fridge} The temperatures of the $^{4}$He bath and of the $^{3}$He evaporator during the operations in preparation of the launch, during the launch and in flight. The zoom in the top panel shows that the temperature drifts are smaller than 0.5~mK/h. In the bottom panel the \emph{red--line} is the time when the motorized valve was closed while the \emph{green--line} is the time when it was opened. 
}
\end{figure}

At 12:30 GMT on July $14^{th}$ the OLIMPO payload reached the float altitude of 37.8~km and both temperatures, of $^{4}$He bath and of the $^{3}$He evaporator, drifted down to 1.65~K and 295~mK, respectively.

\begin{figure}[!ht]
\centering
\includegraphics[width=1\textwidth]{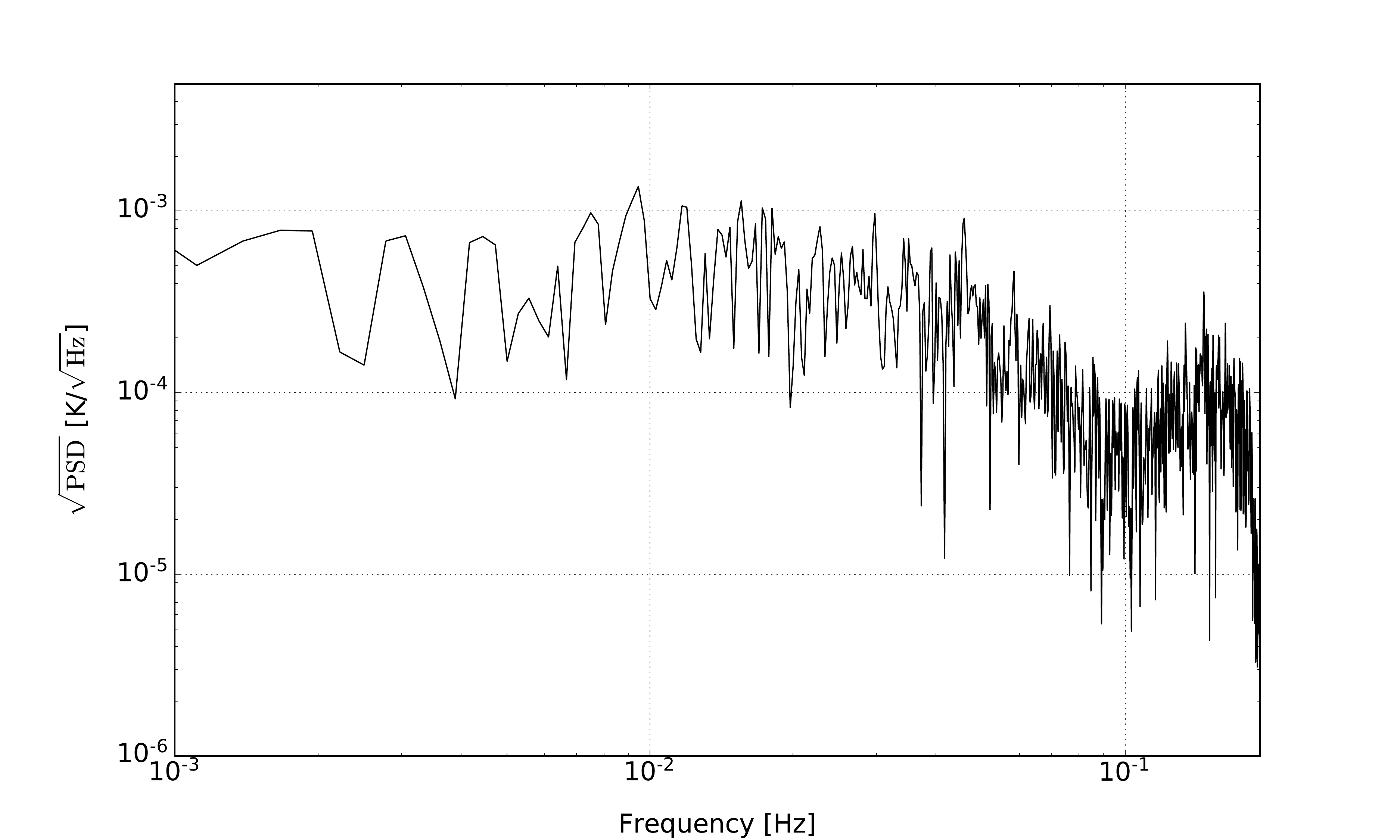}
\caption{\label{psd_fridge_flight} Power spectrum of the temperature fluctuations measured at the 0.3~K $^{3}$He evaporator during one hour of flight data. 
}
\end{figure}

In figure \ref{Temp_Fridge} we plot the temperatures of the $^{4}$He bath and of the $^{3}$He evaporator during  the operations described above. A detailed view of an hour of flight data shows that temperature drifts are smaller than 0.5~mK/h. In figure \ref{psd_fridge_flight} we plot the frequency spectrum of the temperature fluctuations shown in the zoom of figure \ref{Temp_Fridge}. 

Communications with the OLIMPO payload were lost 24 hours after the launch, therefore we do not have temperature records afterwards. However, the temperature of the VCS (as well as those of the other stages) was already stabilized before the connection was lost. Moreover, integrating the evaporation rate computed from the temperature profile of the VCS before communication interruption, we estimate that 33.5~liters of L$^4$He were still present in the tank at the moment of communications black--out. 

At that point the temperature of the VCS was 21~K. With the LN2 tank at 77~K, this temperature would correspond to an evaporation rate of 0.134~l/h. However, at float the LN2 tank sets to a reduced pressure, and its temperature was 52~K, thus reducing the heat load on the VCS with respect to the laboratory tests. This means that an evaporation rate of 0.134~l/h is conservative, since this low temperature of the VCS is also due to the lower thermal load from the LN2 stage. Anyway, assuming this evaporation rate and the estimated volume of L$^4$He remaining in the tank, we estimate a minimum residual hold time in flight of about 11 days. Assuming the same heat load as in the laboratory (which is also a conservative assumption due to the lower radiative load on the $^4$He tank from the shield) we estimate a residual hold time of 14 days.

The payload was separated from the balloon on July 19$^{th}$, 2018, at 02:28 GMT. The cryostat has been subsequently recovered, and the vacuum jacket was still evacuated when arrived in our laboratory 6 months later.

\section{Conclusions}
We have developed a wet cryostat and a self contained $^{3}$He refrigerator, suitable for long hold time (about 15~days) in the stratosphere. The $^{3}$He refrigerator has been optimized to be robust, reliable and suitable for unmanned operation. The cryostat and the $^{3}$He refrigerator have been qualified in a stratospheric balloon flight. The $^4$He tank was sealed before the launch to maintain the L$^4$He pumped during ascent. Once at float, the sealing valve was succesfully re--opened towards the stratospheric vacuum (3~mbar) to resume pumping. After a few hours the system stabilized at the nominal evaporation rate, fullfilling the requirement to cover the nominal duration of the flight (14~days). The $^3$He refrigerator temperature during the flight has been 296~mK with temperature drifts smaller than 0.5~mK/h, providing an excellent environment for the operation of 4 arrays of Kinetic Inductance Detectors \cite{2019JCAP...07..003M, LTD19}.

\section*{Acknowledgements}

We acknowledge the Italian Space Agency (ASI) for funding the OLIMPO experiment, and organizing the 2018 launch campaign.

\section*{References}
\bibliographystyle{alpha}
\bibliography{sample}

\end{document}